\documentclass[twocolumn,pra,draft,nofootinbib,superscriptaddress,showpacs]{revtex4}
\input{psfig}
\usepackage{dcolumn}
\begin{document}
\title{Correlation structure in nondipole photoionization}
\author{M.\ Ya.\ Amusia}
\affiliation{Racah Institute of Physics, The Hebrew University, Jerusalem 91904,
Israel}
\affiliation{A. F. Ioffe Physical- Technical Institute, St. Petersburg 194021, Russia}
\author{A.\ S.\ Baltenkov}
\affiliation{U. A. Arifov Institute of Electronics, Tashkent 700187, Uzbekistan}
\author{ L.\ V.\ Chernysheva}
\affiliation{A. F. Ioffe Physical- Technical Institute, St. Petersburg 194021, Russia}
\author{Z.\ Felfli}
\affiliation{Center for Theoretical Studies of
        Physical Systems, Clark Atlanta University, Atlanta, GA 30314}
\author{S.\ T.\ Manson}
\affiliation{Department of Physics and Astronomy, Georgia State University, Atlanta, GA 30303}
\author{A.\ Z.\ Msezane}
\affiliation{Center for Theoretical Studies of
        Physical Systems, Clark Atlanta University, Atlanta, GA 30314}

\date{\today}
\begin{abstract}
The nondipole parameters that characterize the angular disribution of the photoelectrons 
from the $3d$ subshell of Cs are found to be altered qualitatively by the inclusion of 
correlation in the form of interchannel coupling between the $3d_{3/2}$ and 
$3d_{5/2}$ photoionization channels.  A prominent characteristic maximum is predicted only 
in the parameters for $3d_{5/2}$ photoionization, while the effect for $3d_{3/2}$ is rather 
weak.  The results are obtained within the framework of the Generalized Random Phase 
Approximation with Exchange
(GRPAE), which in addition to the RPAE effects takes into account the rearrangement of all 
atomic electrons due to the creation of a $3d$ vacancy.  
\end{abstract}
\maketitle
Nondipole effects in the photoionization of atoms and molecules were thought to be of any 
importance only at multi-keV photon energies\cite{MD,hand} despite indications to the 
contrary\cite{MOK,cm,balt}.  
However, in an upsurge of work on the subject, it was found both theoretically 
and experimentally that nondipole effects are often of significance at photon energies of 
hundreds and even tens of eV\cite{LH,pratt,cooper,south,jbl,DolMan,oc,non1,jc,he,cher}.  It 
has been determined that photoelectron angular 
distributions are particularly affected by these nondipole photoionization channels.  
In lowest order, the effect of the non-dipole channels is to add two new terms to the 
well-known dipole expression for the photoelectron angular distribution, along with their 
associated dynamical parameters, $\gamma$ and $\delta$, the nondipole 
parameters\cite{balt,pratt,cooper,dolma,peshkin}.  These parameters 
arise from the interference of dipole and quadrupole (E1 and E2) channels in much the same 
manner that the $\beta$ parameter, which characterizes the dipole photoelectron angular distribution, 
arises from dipole-dipole interference\cite{MSrev}.

Studies of these nondipole parameters have found $\gamma$ and $\delta$ to be essentially structureless, 
except in regions of dipole or quadrupole resonances\cite{DolMan,cher} or Cooper 
minima\cite{non1}. It has also been 
averred that correlation effects are unimportant for nondipole effects\cite{AD}.  In this communication, 
we show that assertion to be incorrect.  In fact, we 
report on a new phenomenon, a new kind of structure in nondipole parameters that has nothing 
to do with resonances or Cooper minima.  This structure, as shall be shown below, is due solely to many-body 
correlations; it has no analogue in a single-particle picture.  Furthermore, the structure is 
induced by relativistic interactions; it does not arise in a non-relativistic calculation.  
Thus, the study of this structure is a direct measure of relativistic correlation 
effects.  

To understand the basic origin of this structure, note that it was demonstrated 
recently\cite{prlus} that the interaction between photoionization channels belonging
to different components of the spin-orbit doublets $3d_{3/2}$ and $3d_{5/2}$ in Xe, 
Cs and Ba affect dramatically the partial photoionization cross sections.  Specifically, 
it was shown that due to this interchannel coupling interaction, the partial $3d_{5/2}$ 
cross section acquires an additional prominent maximum.  This result gave the explanation 
for the recent experimental observation of this effect in Xe\cite{kiv}  and also 
predicted similar, even more dramatic effects for $3d$ photoionization in Cs and Ba.

In this communication we present the results of our studies, based on the 
approach developed earlier\cite{prlus},
of the nondipole angular distribution parameters for $3d$ 
photoionization in Cs to illustrate a new phenomenon where
impressive manifestations of intra-doublet interchannel interactions 
producing new structures in nondipole parameters are predicted, thereby also contradicting 
the notion that correlation does not affect nondipole parameters significantly.

The theoretical results were obtained using a specially modified random-phase-approximation 
with exchange (RPAE) approach
previously developed for half-filled subshells\cite{amujpb,amusov}.  This calculation
showed clearly the physics of the phenomenon observed\cite{kiv} and led to results\cite{prlus} 
that are in both qaulitative and quantitative agreement with experimental 
$3d$ Xe data.  Explicitly
relativistic calculations\cite{rado} were subsequently performed using the 
relativistic-random-phase approximation (RRPA)\cite{johnson} 
which confirmed the qualitative and quantitative 
accuracy of the results in \cite{prlus}.  

To summarize the theoretical approach, we consider 
the $3d_{5/2}$ and $3d_{3/2}$ subshells each as half-filled atomic subshells.  This permits 
us to apply straightforwardly the RPAE methodology to include many-body correlations 
(including interchannel coupling)
for half-filled subshells.  Exchange is neglected between these two sorts of electrons,
the six that form the $3d_{5/2}$ (called "up"), and the four forming $3d_{3/2}$
(called "down") electrons.  However, in the real half-filled 3d subshell one would 
have five electrons.  But these corrections, 6/5 and 4/5, respectively can be introduced
easily into the calculational scheme.

Then we concentrate on the investigation of the influence of "up" and "down" electrons
upon each other and demonstrate that the effect of the $3d_{3/2}$ ("down") electrons 
upon the $3d_{5/2}$ ("up")
manifests itself not only in the partial cross sections $\sigma_{5/2}(\omega)$, 
$\sigma_{3/2}(\omega)$\cite{prlus} and somewhat in the respective angular 
anisotropy parameters $\beta_{5/2}(\omega)$,
$\beta_{3/2}(\omega)$\cite{rado}, but also in the nondipole parameters of the photoelectron
angular distributions, $\gamma(\omega)$ and $\delta(\omega)$.  In all these cases 
the effect of the $3d_{3/2}$ photoionization channels upon the $3d_{5/2}$ leads 
to the creation of an additional
maximum, while the action of the $3d_{5/2}$ electrons upon $3d_{3/2}$ proved to 
be generally negligible.
Note that since the matrix elements for the photoionization process are strongly 
modified by correlation in the form of interchannel coupling, it is evident that
characteristics of the process other than the integrated cross section; for example, 
the photoelectron spin polarization
(see \cite{cherepkov} and references therein) are also modified.

The angular distribution of photoelectrons from an $nl$ subshell created by linearly polarized 
light, including the lowest order nondipole parameters\cite{balt,pratt,cooper,dolma,peshkin} 
is given by

\begin{eqnarray}
\frac{d\sigma _{nl}(\omega )}{d\Omega }&=&\frac{\sigma _{nl}(\omega )}{4\pi }
[1+{\beta _{nl}(\omega )}P_{2}(\cos\theta)\nonumber \\
&+&[\delta_{nl}(\omega)+\gamma_{nl}(\omega)cos^{2}\theta]
sin\theta cos\phi ],
\end{eqnarray}
where $\sigma _{nl}(\omega )$ is the partial cross section, $\beta _{nl}(\omega)$ 
is the dipole angular anisotropy parameter, $\gamma_{nl}(\omega)$ and $\delta _{nl}(\omega)$ 
are nondipole parameters, $P_{2}(\cos\theta)$ is a Legendre polynomial, $\theta$ is the 
angle between the photoelectron and the photon polarization directions, and $\phi$
is the angle between the photon momentum and the projection of the photoelectron momentum 
in the plane perpendicular to the photon polarization.
The expressions for $\gamma_{nl}(\omega)$ and $\delta _{nl}(\omega)$ in terms of
dipole and quadrupole matrix elements were first obtained in a slightly different form 
in \cite{balt} and later in \cite{pratt,cooper,non1,non2}.
Most of the attention in the past was given to $l=0$ and $l=1$, 
where the expressions for $\gamma_{nl}$ 
and $\delta_{nl}$ are relatively simple. Here we are interested in the more complex case 
of $l=2$, and the explicit expressions are given elsewhere\cite{cooper}.  For 
present purposes, however, it is necessary to point out that the expressions involve 
ratios of quadrupole to dipole matrix elements, along with cosines of phase shift 
differences.

In the present calculation, the $3d \rightarrow \epsilon p,\epsilon f$ dipole 
amplitudes and $3d \rightarrow \epsilon s,\epsilon d,\epsilon g$ quadrupole 
amplitudes are taken into account for both $3d_{5/2}$ and $3d_{3/2}$ photoionization.
The matrix elements for dipole, $d_{l\pm 1}$, and quadrupole, $q_{l\pm 2,0}$, are defined as
\begin{eqnarray}
d_{l\pm 1} \equiv d_{nl,\epsilon l\pm 1} = \int\limits_{0}^{\infty} \phi_{nl}(r) r \phi_{\epsilon l \pm 1}(r) dr, \nonumber \\
q_{l\pm 2,0} \equiv q_{nl,\epsilon l\pm 2,0} = \frac{1}{2} \int \limits_{0}^{\infty} \phi_{nl}(r) r^{2} \phi_{\epsilon l \pm 2,0}(r) dr. 
\end{eqnarray}
Here $\phi_{nl}(r)$ and $\phi_{\epsilon l^{\prime}}(r)$ are the radial parts of the 
Hartree-Fock (HF) one-electron wave functions. In the uncorrelated (HF) approximation, 
the formulae for $\gamma_{nl}(\omega)$ and 
$\delta _{nl}(\omega)$ can be used directly with the dipole and quadrupole 
amplitudes.

An effect of correlation, however, is to render these transition matrix 
elements complex; the correlated dipole amplitudes are termed $D_{l\pm 1}$ and 
the correlated quadrupole amplitudes $Q_{l\pm 2,0}$.  To obtain the corresponding 
expressions for $\gamma_{nl}$ and $\delta_{nl}$, one has to perform the following 
substitutions\cite{balt,dolma} 

\begin{eqnarray}
&\mid d_{l\pm 1} \mid ^{2} \to \mid Re  D_{l\pm 1} \mid ^{2} + \mid Im  D_{l\pm 1} \mid ^{2} \nonumber \\
&\mid q_{l\pm 2,0} \mid ^{2} \to \mid Re  Q_{l\pm 2,0} \mid ^{2} + \mid Im  Q_{l\pm 2,0} \mid ^{2} \nonumber \\
&d_{l\pm 1} q_{l\pm 2,0} \cos(\delta_{l\pm 2,0} - \delta_{l\pm 1} ) \to [ Re D_{l\pm 1} Re Q_{l\pm 2,0} \nonumber \\ 
&+ Im  D_{l\pm 1} Im  Q_{l\pm 2,0})] \cos(\delta_{l\pm 2,0} -\delta_{l\pm 1} ) - \nonumber \\
&- [ Re D_{l\pm 1}  Im Q_{l\pm 2,0} - Im D_{l\pm 1} Re Q_{l\pm 2,0} ) ]  \nonumber \\
&\sin(\delta_{l\pm 2,0} - \delta_{l\pm 1} ) 
\end{eqnarray}

\noindent where the $\delta _{l'}$ are the single-particle (HF) phase shifts for the 
designated channel, and $Re$ and $Im$ stand for real and imaginary 
parts, respectively. 
 
The method that we use here, just as in \cite{prlus}, is the spin-polarized random-phase 
approximation with exchange (SPRPAE).  However, for the intermediate $3d$ subshell SPRPAE 
is not sufficient; the effects of core rearrangement (relaxation) must be taken into account.  
This is done by going from RPAE to the generalized (GRPAE) or, in our case, from SPRPAE to 
SPGRPAE which takes into account that while a slow photoelectron leaves the atom, the field 
seen is modified due to the alteration (relaxation) of the wave functions of all other atomic 
electrons as a result of the creation of the inner-subshell vacancy.  GRPAE is discussed at 
length in \cite{book}.  Its extension for a system with two types of electrons "up" and "down", 
the transition from SPRPAE to SPGRPAE, is straightforward.

The results of our calculations for $\gamma$ for photoionization of Cs $3d_{5/2}$ and 
$3d_{3/2}$ are shown in Fig. 1; both the correlated and the uncorrelated Hartree-Fock 
(HF) results are shown to emphasize the 
features brought about by correlation.  As seen from the figure, the outstanding 
difference between correlated and uncorrelated results appears in the value of
$\gamma$ for $3d_{5/2}$ photoionization in the region just above a photon 
energy of 740 eV.  The fact that this significant structure in $\gamma$ is not seen in 
the uncorrelated, HF, calculation is proof that its existence is due to 
correlation.  Specifically, we have found that it is due to interchannel coupling 
among the $3d_{5/2}$ and $3d_{3/2}$ photoionization channels; by interchannel 
coupling we mean simply configuration interaction in the continnum.  Since  
the "real" wave function for the final continuum state in a photoionization process 
is multichannel, there is mixed in with the final 
state wave functions for $3d_{5/2}$ photoionization a small amount of $3d_{3/2}$ channels.  
Then, due to the $3d \rightarrow \epsilon f$ shape resonance in the region just above
the $3d_{3/2}$ threshold, the $3d_{3/2}$ dipole cross section is much larger than the 
$3d_{5/2}$.  Thus mixing a small amount of the wave function of the 
$3d_{3/2} \rightarrow \epsilon f$ channel with the wave function of the much smaller $3d_{5/2}$ 
channel alters the $3d_{5/2}$ dipole matrix elements significantly; the $3d_{5/2}$ 
cross section maximizes with the $3d_{5/2} \rightarrow \epsilon f$ just above its threshold,
and drops off considerably at the energy of the $3d_{3/2} \rightarrow \epsilon f$ shape 
resonance.  

But the $\gamma$ parameter, as discussed above, is essentially a ratio of quadrupole to 
dipole matrix elements.  There is also interchannel coupling among the quadrupole channels,
$3d_{5/2} \rightarrow \epsilon s,d,f$ and $3d_{3/2} \rightarrow \epsilon s,d,f$.  But there 
are no significant resonances in any of these channels in the energy range of interest, and 
all of the channels are of more or less the same size.  Thus, interchannel coupling does 
not introduce any important changes in the quadrupole photoionization matrix elements like 
it does in the dipole case.  Since only one participant in the ratio that yields $\gamma$ 
changes significantly because of correlation, then it is evident that the ratio itself must 
be altered as well.  This is the reason for the significant structure seen in Fig. 1 for
$\gamma$ of $3d_{5/2}$ in the vicinity of the $3d_{3/2}$ threshold.

The only other real effect of correlation on the $\gamma$ parameter is for $3d_{3/2}$
right at threshold, where the HF result is noticeably lower than the correlated value.
This again is due to interchannel coupling, but in this case the effect of the $3d_{5/2}$ 
photoionization channels on the $3d_{3/2}$ causes the modification.  In particular, the 
$3d_{3/2} \rightarrow \epsilon f$ amplitude, which is extremely small at the $3d_{3/2}$
threshold, is strongly modified by its interchannel interaction with the much larger
$3d_{5/2}$ amplitudes.

The situation for the $\delta$ parameter is shown in Fig. 2 and we find here a story 
similar to that of the $\gamma$ parameter.  A significant correlation structure in 
the $3d_{5/2}$ channel, seen just above the $3d_{3/2}$ threshold is evident, 
along with changes in $\delta$ in the $3d_{3/2}$ channel just at its threshold.  These 
effects of correlation occur for exactly the same reason as discussed above for $\gamma$.  
It is apparent from comparison of Figs. 1 and 2 that the changes in $\delta$ due to 
correlation are qualitatively larger than for $\gamma$.  This is probably due to the 
fact that the values of $\delta$ are roughly an order of magnitude smaller than $\gamma$.

Owing to the geometry of a number of experimental setups to measure nondipole photoelectron 
angular distributions, $\gamma$ and $\delta$ are often not measured individually\cite{south,jbl}; 
instead, $\zeta$ = $\gamma$ + 3$\delta$ is investigated.  To connect with such experimental 
investigations, $\zeta$ for the $3d_{5/2}$ and $3d_{3/2}$ channels of Cs are shown in Fig. 3.  
Since $\zeta$ is simply a linear combination of $\gamma$ and $\delta$, the explanation for 
the correlation structure here simply follows from the previous discussion for $\gamma$ 
and $\delta$ individually.

In conclusion then, it has been shown that, due to spin-orbit induced 
interchannel coupling between the 
dipole photoionization channels arising from the two members of the $3d$ spin-orbit 
doublet in Cs, a new structure results in the nondipole photoelectron angular distribution 
parameters $\gamma$ and $\delta$ (and, of course, $\zeta$).  Similar results are found 
for Xe $3d$ and Ba $3d$; these will be presented in a future publication.  This is the 
first case of structure in the energy dependence of nondipole parameters discovered that 
is not related to resonances (dipole or quadrupole) or Cooper minima.  And, it was shown
explicitly that correlation in the form of interchannel coupling can indeed be important 
for the nondipole photoelectron angular distribution parameters, in contradistinction 
to the conventional wisdom.  Experimental scrutiny 
into this prediction is highly desirable.

Research at Clark Atlanta University is supported by DoE, Division of Chemical Sciences,
Office of Basic Energy Sciences, Office of Energy Research (AZM) and NSF.
The work of MYaA was supported by the Hebrew University Intramural Fund.  MYaA and LVC 
acknowledge the support of the International Science and Technology Center (grant No 1358).
The work of ASB  and STM was supported by the US Civilian Research $\&$
Development Foundation for the Independent States of the Former Soviet Union, 
Award N0. ZP1-2449-TA-02. STM also acknowledges the support of NSF and NASA.

\begin{center}
\section*{Figure Captions}
\end{center}
\begin{enumerate}
\item[FIG. 1] Calculated values of the nondipole photoelectron angular distribution parameter 
$\gamma$ for Cs $3d_{5/2}$ and $3d_{3/2}$ subshells in correlated (SRPAE) and uncorrelated 
(HF) approximations.
\item[FIG. 2] Calculated values of the nondipole photoelectron angular distribution parameter 
$\delta$ for Cs $3d_{5/2}$ and $3d_{3/2}$ subshells in correlated (SRPAE) and uncorrelated 
(HF) approximations.
\item[FIG. 3] Calculated values of the nondipole photoelectron angular distribution parameter 
$\zeta$ for Cs $3d_{5/2}$ and $3d_{3/2}$ subshells in correlated (SRPAE) and uncorrelated 
(HF) approximations.

\end{enumerate}
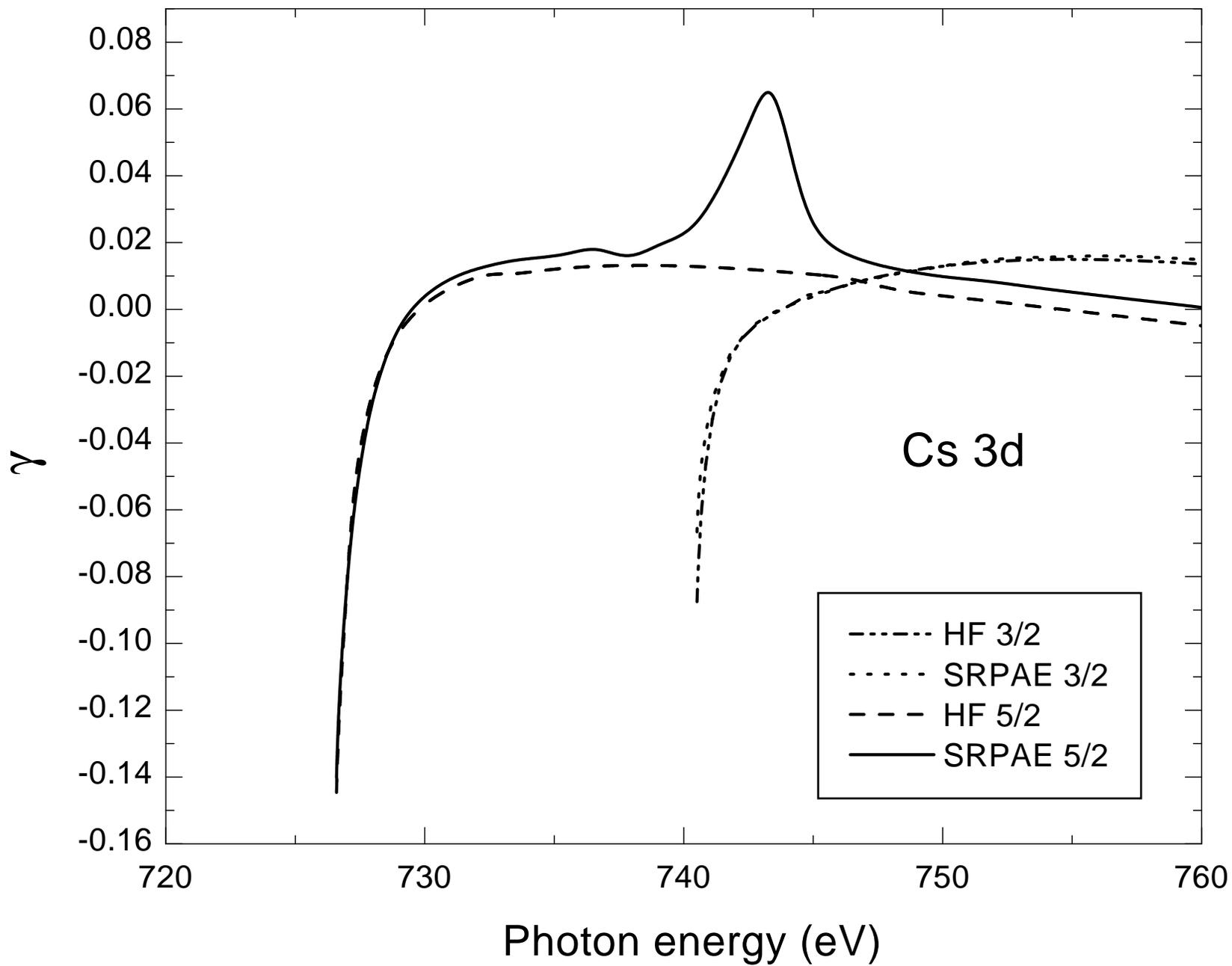
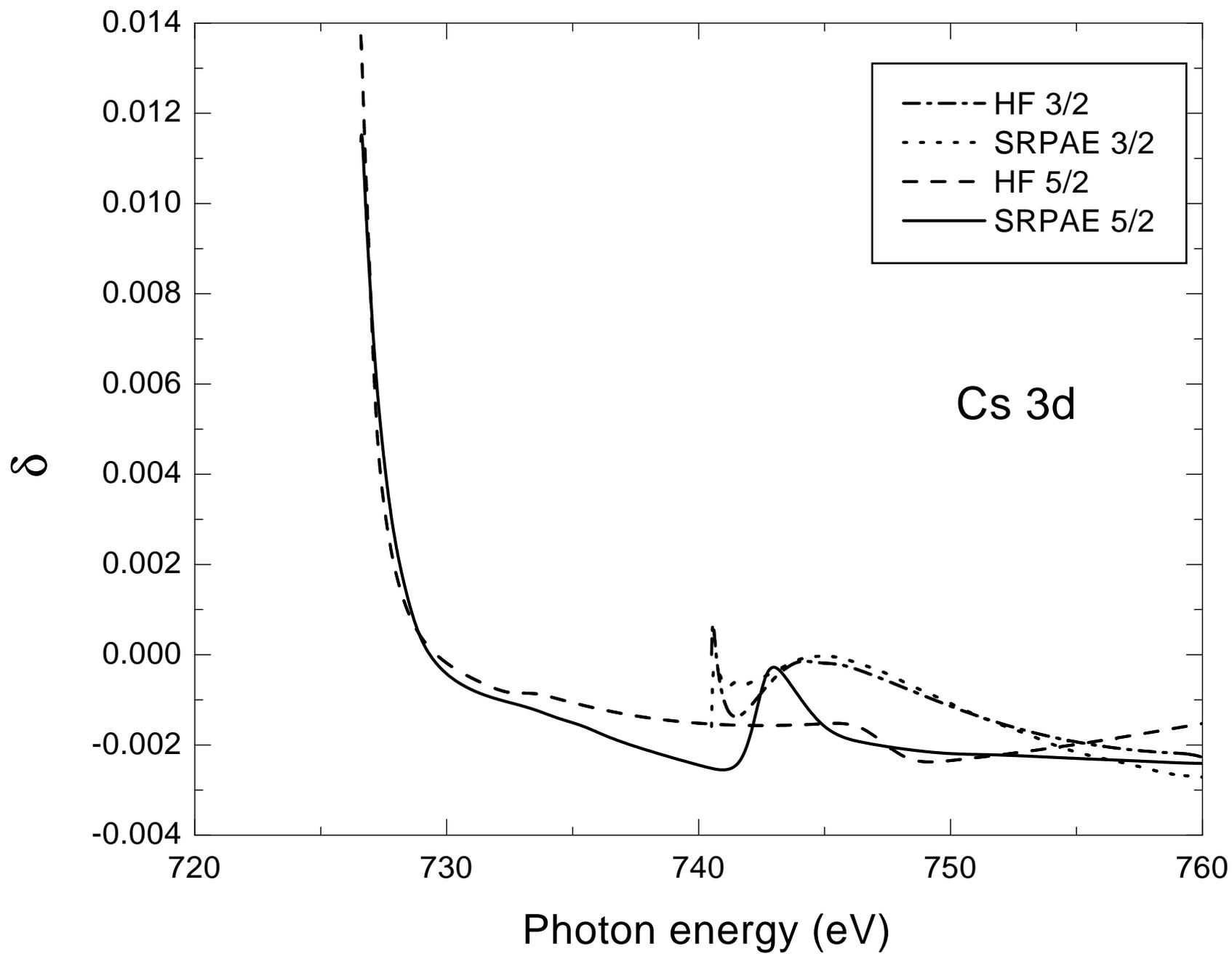
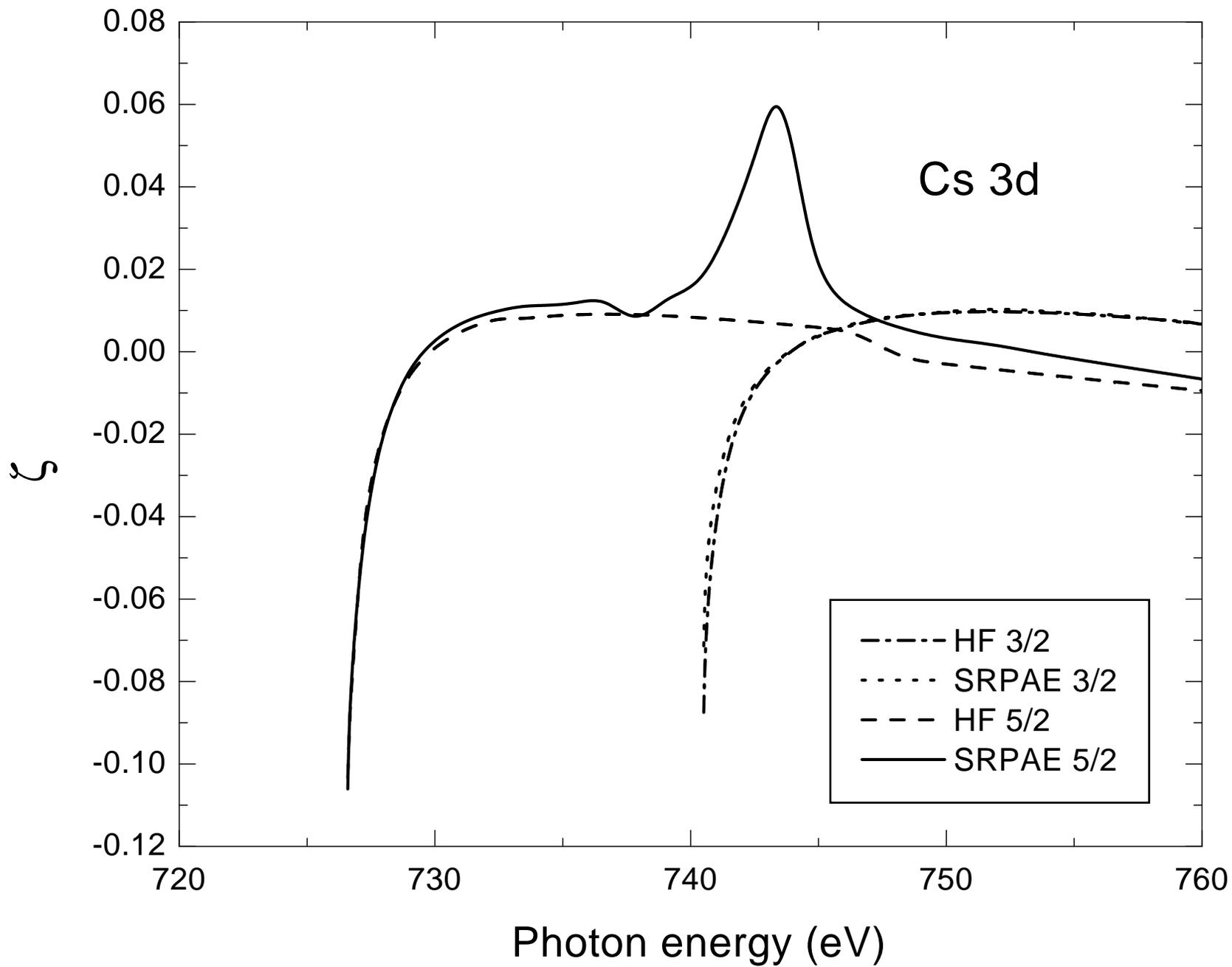
\end{document}